# High resolution spectroscopy of methyltrioxorhenium: towards the observation of parity violation in chiral molecules[†]

Clara Stoeffler[1], Benoît Darquié[1,*], Alexander Shelkovnikov[1], Christophe Daussy[1], Anne Amy-Klein[1], Christian Chardonnet[1], Laure Guy[2], Jeanne Crassous[3], Thérèse R. Huet[4], Pascale Soulard[5], Pierre Asselin[5,*]

[1] *Laboratoire de Physique des Lasers, UMR7538 Université Paris 13-CNRS, 99 av. J.-B. Clément, F-93430 Villetaneuse, France*
[2] *Laboratoire de Chimie, UMR 5182 Ecole Normale Supérieure de Lyon-CNRS, 46 Allée d'Italie, F-69364 Lyon 07, France*
[3] *Sciences Chimiques de Rennes, Campus de Beaulieu, UMR6226 Université de Rennes 1-CNRS, 35042 Rennes Cedex, France*
[4] *Laboratoire de Physique des Lasers, Atomes et Molécules, UMR8523 Université Lille 1-CNRS, F-59655 Villeneuve d'Ascq Cedex, France*
[5] *Laboratoire de Dynamique, Interactions et Réactivité, UMR7075 Université Pierre et Marie Curie-CNRS, 4 Place Jussieu, F-75005 Paris, France*
[*] *to whom correspondence should be addressed:* `benoit.darquie@univ-paris13.fr`, `pierre.asselin@upmc.fr`

## Abstract

Originating from the weak interaction, parity violation in chiral molecules has been considered as a possible origin of the biohomochirality. It was predicted in 1974 but has never been observed so far. Parity violation should lead to a very tiny frequency difference in the rovibrational spectra of the enantiomers of a chiral molecule. We have proposed to observe this predicted frequency difference using the two photon Ramsey fringes technique on a supersonic beam. Promising candidates for this experiment are chiral oxorhenium complexes, which present a large effect, can be synthesized in large quantity and enantiopure form, and can be seeded in a molecular beam. As a first step towards our objective, a detailed spectroscopic study of methyltrioxorhenium (MTO) has been undertaken. It is an ideal test molecule as the achiral parent molecule of chiral candidates for the parity violation experiment. For the $^{187}$Re MTO isotopologue, a combined analysis of Fourier transform microwave and infrared spectra as well as ultra-high resolution $CO_2$ laser absorption spectra enabled the assignment of 28 rotational lines and 71 rovibrational lines, some of them with a resolved hyperfine structure. A set of spectroscopic parameters in the ground and first excited state, including hyperfine structure constants, was obtained for the $v_{as}$ antisymmetric Re=O stretching mode of this molecule. This result validates the experimental approach to be followed once a chiral derivative of MTO will be synthesized, and shows the benefit of the combination of several spectroscopic techniques in different spectral regions, with different set-ups and resolutions. First high resolution spectra of jet-cooled MTO, obtained on the set-up being developed for the observation of molecular parity violation, are shown, which constitutes a major step towards the targeted objective.

---

[†] Supplementary information available in the ancillary file: Measured and fitted transition frequencies as well as spectroscopic constants.

# I. Introduction: context and history

Chiral molecules have been identified as relevant candidates for the study of parity violating effects arising from the exchange of virtual $Z_0$ bosons between electrons and constituents of the nuclei, a mechanism precisely described by standard electroweak theory. It should lead to small energy differences between the two enantiomeric mirror image states of chiral molecules[1-3], but to date, no experiment has reached the sensitivity required to observe this tiny difference. The weakness of the effect represents a very difficult experimental challenge. There has been speculation that this minute difference could still be large enough to trigger the observed homochirality in living organisms, the biochemistry showing, with very few exceptions, a distinct *preference* for *L*-amino acids and *D*-monosaccharides over their mirror images. The role of weak interaction in the origin of homochiral life is still a subject of debate[4-10], which is one motivation for an experimental observation of parity violation (PV) in molecules that would enable to address this question quantitatively.

A number of experimental techniques have been proposed for the observation of PV in molecular systems, including rovibrational[11], electronic[12-14], Mössbauer[15] and NMR[16] spectroscopy, as well as crystallization[17] and solubility[18] experiments, or optical activity measurements[19], but no unambiguous observation has so far been made. After the pioneering work of Yamagata[1], Rein[2], and Gajzago and Marx[3] who first suggested that weak interaction is responsible for an energy difference in the spectrum of right- and left-handed chiral molecules, Letokhov proposed in 1975 to observe PV effects in molecules as a difference $\Delta \nu_{PV} = \nu_L - \nu_R$ in the rovibrational frequencies $\nu_L$ and $\nu_R$ of the same transition of left and right enantiomers.[11] After a first test on separated enantiomers of camphor in 1977[20], the most sensitive experiments to date following Letokhov's proposal were performed around 2000 by authors of this paper.[21, 22] A $CO_2$ laser-based spectrometer was developed to probe a bromochlorofluoromethane (CHFClBr) transition around $\nu \sim 30$ THz (10 µm or 1000 cm$^{-1}$), using saturated absorption laser spectroscopy. The spectrum of each enantiomer was simultaneously recorded in separate Fabry-Perot cavities, and the absorption frequencies were compared at a $5 \times 10^{-14}$ level. However, a residual pressure shift induced by impurities in the samples was observed and found to limit the experimental sensitivity at $2 \times 10^{-13}$ for the parity violation effect ($\nu_L - \nu_R < 8$ Hz), taking into account the respective enantiomeric excess of 56% and 72% for the *R*-(–) and *S*-(+) configuration samples. Later theoretical studies predicted a PV difference for CHFClBr on the order of -2.4 mHz[23, 24], corresponding to $\Delta \nu_{PV} / \nu \approx -8 \times 10^{-17}$. To go further, it was necessary to develop a new experiment in which, among other improvements, collisional effects would be negligible.

A new generation of $CO_2$ laser spectroscopy experiments based on the recording of two-photon Ramsey fringes of chiral molecules in a supersonic jet is under development at Laboratoire de Physique des Lasers (LPL).[25-28] The experiment is adapted from an existing Ramsey fringes set-up that enabled us to measure $SF_6$ absolute frequencies with a sub-Hz ($10^{-14}$) accuracy.[29, 30] In particular, it will benefit from the accurate control of the $CO_2$ laser frequency by comparison to a primary frequency standard. Besides we expect a strong reduction of systematic effects for three reasons: (i) in such a supersonic jet set-up, collisions are mainly due to background pressure, 4 to 5 orders of magnitude lower than in the CHFClBr experiment, and the corresponding shift will be reduced accordingly; (ii) the same set-up will be alternatively fed with right- and left-handed molecules which will cancel out a number of systematic effects; (iii) the probed line width will be of the order of 100 Hz, instead of 60 kHz for the CHFClBr experiment, which will reduce most systematic errors accordingly. With this set-up, a sensitivity of ~0.01 Hz ($3\times10^{-16}$) on the frequency difference between two enantiomers seems accessible. For more details, the reader is advised to refer to [25].

The expected PV effect scales approximately as $Z_A^5$ ($Z_A$ is the nuclear charge of the most heavy nucleus).[31, 32] Thus candidate molecules should preferably contain one or more heavy atoms at or near the chiral centre. Lately, chiral selenium, tungsten, gold, mercury, iridium, osmium and rhenium complexes have been calculated, by P. Schwerdtfeger, R. Bast and coworkers, to be favourable candidates for PV observation.[33-39] For instance, the rhenium complex Re($\eta^5$-Cp*)(=O)(CH$_3$)Cl (**1**) (Figure 1) presents a strong vibrational band in the 10 µm spectral region of the CO$_2$ laser, and a large PV effect $\Delta\nu_{PV}$ of 2.39 Hz (~$10^{-13}$)[35] (obtained from 4-component relativistic Hartree-Fock calculations) which we should detect unambiguously with the new generation set-up.

The sensitivity of our experiment aiming at observing PV effects in molecules will highly depend on the Ramsey fringes signal-to-noise ratio, the signal being proportional to the molecular flux. Even if it is conceivable for synthetic chemists to produce chiral-at-metal complexes similar as **1** in large quantity and with an enantiomeric excess of 100%, these molecules are solid at room temperature, which constitutes a major difficulty for high resolution molecular beam gas phase spectroscopy. Since methyltrioxorhenium (MTO) is an oxorhenium complex known to sublimate easily, an interesting strategy is to study chiral MTO derivatives. Parallel to the current development of appropriate chiral MTO derivatives for PV observation, a detailed spectroscopic study of MTO has been undertaken and is presented in this article. This work enabled us to validate our experimental approach and develop the set-up dedicated to the PV observation. In section II, we briefly review the families of oxorhenium complexes that were synthesized for our new PV test, and explain why we foresee favourable molecules from achiral MTO. Section III describes a Fourier Transform Microwave (FTMW) spectroscopy study of MTO and section IV reports on the first rovibrational spectroscopic characterization of the $\nu_{as}$ antisymmetric Re=O stretching mode of this molecule via Fourier Transform Infra Red (FTIR) spectroscopy. Room temperature and jet-cooled spectra have been recorded. A list of observed lines were assigned in $J$ and $K$ (and $F$ for rotational transitions) and a list of calculated transition frequencies has been proposed bringing to light a number of lines accessible to the CO$_2$ laser. In section V, ultra-high resolution CO$_2$ laser saturated absorption spectroscopy of room temperature MTO in a cell is described, giving credit to the previous analysis. A first relevant assignment of fully resolved hyperfine transitions is proposed. First high resolution spectra of jet-cooled MTO, obtained on the set-up dedicated to molecular PV effects observation, are shown, which constitutes a major step towards the targeted test.

## II. Search of candidate molecules

The choice of the molecule is presently the most challenging issue since it needs to fulfil a series of stringent conditions. A chiral molecule should have a strong absorption band in the mid-infrared and a large PV effect in the rovibrational spectrum (*i.e.* $\Delta\nu_{PV}$ ~ 1 Hz). It must be available in the two enantiopure forms and in large quantity (a few grams). It must sublimate without decomposition in order to enable the production of a supersonic molecular jet. In addition, we will prefer the simplest molecules with, if possible, no low-frequency vibrational or internal rotation mode and the smallest possible number of hyperfine sublevels. These are the conditions needed to obtain a favourable partition function, which has a direct impact on the signal-to-noise ratio and thus the sensitivity of the experiment, since we probe a single quantum level.

Following the relativistic calculations of Schwerdtfeger and Bast[35], we have considered chiral oxorhenium complexes (**1**, Figure 1) as appropriate molecules for a possible experimental observation of PV in chiral molecules. Indeed, these molecules fulfil most of the above requirements.

Since 2006, we have focused on different classes of compounds, *i.e.* chiral oxorhenium complexes such as **2** bearing a hydrotris(1-pyrazolyl)borate (Tp) and a chiral amino-alcohol as ligands, or "3+1 mixed" oxorhenium complexes such as **3**, based on sulfurated ligands (Figure 1).[25] PV shifts of the order of 100 mHz have been calculated for some of the sulfurated Re complexes.[25, 40] Their synthesis in enantiopure form has been studied. Diverse structural aspects have been investigated, among which their stereochemistry and conformational analysis. High Performance Liquid Chromatography (HPLC) separations over chiral stationary phases of these neutral complexes have revealed themselves efficient for the preparation of pure enantiomers [40-42]. However the sublimation of all these complexes appeared difficult, preventing any possibility to efficiently prepare a molecular jet.[25]

To circumvent this problem, we recently turned to MTO (($CH_3$)$ReO_3$, **4** on Figure 1) a molecule commercially available, known to sublimate very efficiently and at relatively low temperature.[25, 43] Therefore, our new target molecules are now chiral derivatives of MTO such as molecule **5** having a tetrahedral chiral rhenium centre. A PV vibrational frequency difference of 400 mHz ($\sim 10^{-14}$) has been recently estimated for this compound by T. Saue and co-workers.[25, 44] This molecule is thus an ideal prototype for the search of molecular PV. As a first step we decided to study achiral MTO, the parent molecule of envisaged candidates for the PV experiment. It is an ideal test molecule which will enable us to validate our experimental approach and calibrate our set-up, via the study of the antisymmetric Re=O stretching mode, that will exist in chiral derivatives even if the overall rovibrational spectrum will be different.

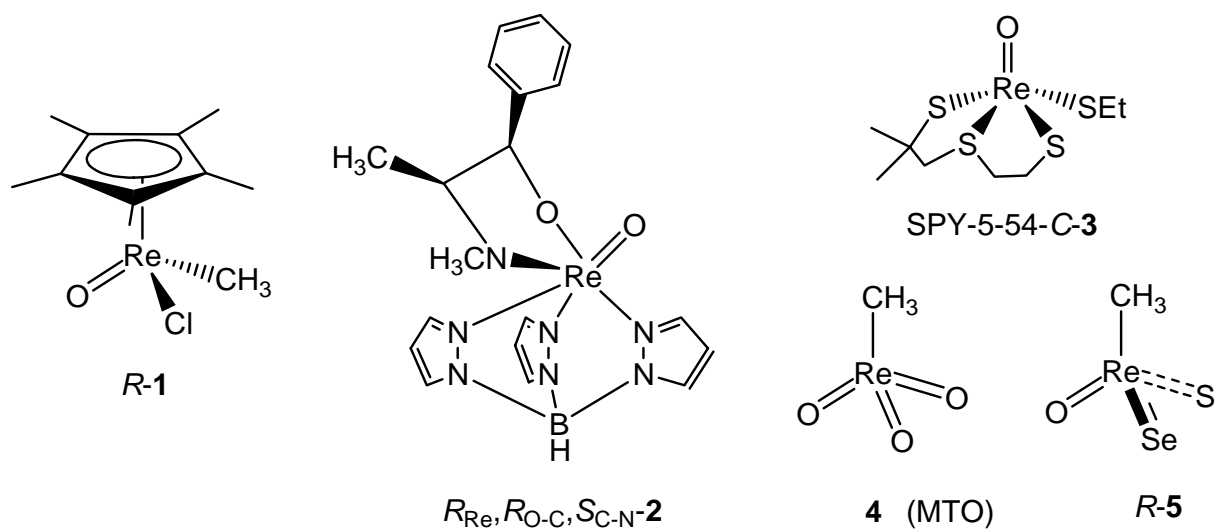

**Figure 1:** Classes of chiral and achiral oxorhenium complexes that have been considered for the experiment.

Once a candidate molecule is identified, there is most of the time very little spectroscopic data available. This is the reason why our work involves experts in various kinds of spectroscopies in order to acquire step-by-step a precise knowledge on spectral line positions. A typical 10 MHz accuracy on the position of a rovibrational line is needed in order to be studied with the ultra-high resolution spectrometer at LPL. For this purpose, high resolution rotational spectroscopy (section III) and rovibrational spectroscopy (section IV) of MTO were examined.

MTO powder (from Strem Chemicals Inc., 98 % purity) has been used without purification in all the experiments described in this paper.

## III. Microwave rotational spectroscopy of MTO

The ground-state rotational constants of the two $^{187}$Re and $^{185}$Re isotopologues of MTO have previously been determined using microwave spectroscopy by Sickafoose *et al*.[45] MTO rotational spectroscopy has recently been re-investigated in T. Huet's group (at Laboratoire de Physique des Lasers, Atomes et Molécules, PhLAM) with a more sensitive spectrometer in the 2-20 GHz range. The main characteristics of the spectrometer consists of two large mirrors of diameter of 0.7 m, in order to maximise the signal-to-noise ratio at low frequencies (diffraction losses), and an high signal acquisition repetition rate of 120 MHz.[46] The MTO crystalline powder was heated at about 320 K and the vapour was mixed with neon carrier gas at a backing pressure of 1.5 bar. The mixture was introduced into a Fabry-Perot cavity at a repetition rate of 1.5 Hz. Molecules were polarized within the supersonic expansion by a 2 µs microwave pulse and the free induction decay signal was detected and digitized at a repetition rate of 120 MHz. After transformation of the signal from the time to the frequency domain, molecular lines were observed as Doppler doublets, with a point every 0.46 kHz, resulting from the average of about 100 signals. The transition frequency was deduced from the average of the two Doppler components with an uncertainty estimated to be 1 kHz for most of the lines. Compared to the previous work[45], all the twenty-eight hyperfine lines associated with the $J' \leftarrow J \equiv 1 \leftarrow 0$ and $2 \leftarrow 1$ rotational transitions were recorded, increasing by a factor of two the number of reported lines. Consequently a better uncertainty could be obtained on several molecular vibrational ground state parameters. This new set of values is reported in Table I (together with the previous set from Sickafoose *et al*) and will be used for the present rovibrational analysis of the Re=O antisymmetric stretching vibrational mode (section IV). The molecular parameters considered were the $B$ rotational constant, the $D_J$ and $D_{JK}$ quartic centrifugal distortion parameters, the $eQq$ quadrupole coupling constant associated with the Re nuclear hyperfine quadrupole interaction, and the $C_{aa}$ and $C_{bb}$ spin-rotation parameters of the Re nucleus (because of the C$_{3v}$ symmetry, $C_{bb}=C_{cc}$).[45] This set of parameters was obtained for both the $^{185}$Re and $^{187}$Re MTO isotopologues using the SPFIT program (developed by H. Pickett, Jet Propulsion Laboratory[47]), giving a root-mean-square deviation on the difference between the observed and calculated transition frequencies (obs-cal) of 2.7 kHz, for both species. The list of assigned lines is available in the ancillary file.

**Table I :** Molecular parameters of CH$_3$$^{187}$ReO$_3$ and CH$_3$$^{185}$ReO$_3$ in the ground state, determined from microwave data. The parameter set from the pioneering microwave study of Sickafoose *et al*[45] is reported for comparison. The listed uncertainties are 1σ.

|  | CH$_3$$^{187}$ReO$_3$ | | CH$_3$$^{185}$ReO$_3$ | |
| --- | --- | --- | --- | --- |
|  | Ref. [45] | This work | Ref. [45] | This work |
| $A$ (MHz) | - | - | - | - |

| $B$ (MHz) | 3466.964(2) | 3466.96481(39) | 3467.049(3) | 3467.04957(39) |
| --- | --- | --- | --- | --- |
| $D_J$ (kHz) | 0.7(2) | 0.705(50) | 0.6(4) | 0.755(50) |
| $D_{JK}$ (kHz) | 1.9(1.0) | 2.208(118) | 2.1(1.4) | 1.971(118) |
| $eQq$ (MHz) | 716.546(17) | 716.54005(192) | 757.187(25) | 757.18175(192) |
| $C_{aa}$ (kHz) | -50(8) | -52.22(37) | -45(11) | -50.66(37) |
| $C_{bb}$ (kHz) | -51.7(6) | -51.464(92) | -51.8(1.1) | -51.165(92) |

Owing to the $C_{3v}$ symmetry of MTO, the $A$ rotational constant cannot be determined by microwave spectroscopy and we are not aware of any reliable gas phase value in the literature. However, in reference[48], Wikrent *et al* recorded rotational spectra of six isotopologues of MTO and were able to determine a three-dimensional structure. Starting from this geometry and imposing the $C_{3v}$ symmetry constraints, we calculated a value for $A$ of 3849.81 MHz with an estimated uncertainty of the order of 10 MHz.

## IV. FTIR spectroscopy of MTO

MTO in the gas phase is a prolate symmetric top with $C_{3v}$ symmetry and 18 vibrational modes.[49] In the 1000 cm$^{-1}$ region two stretching modes are infrared active: the $\nu_{as}(E)$ and $\nu_s(A_1)$ bands which correspond to the antisymmetric and symmetric vibrations of the Re=O bond, whose band centres were measured at 959 and 998 cm$^{-1}$ respectively in the solid phase at room temperature.[50]

### A. Experimental results

Room temperature FTIR spectra have been recorded in an 85-cm static stainless steel cell (described elsewhere[51]) at different instrumental resolutions ranging from 0.006 to 0.1 cm$^{-1}$. Cold fundamental Re=O stretching and related hot bands of MTO could be observed with the best signal-to-noise ratio, at 0.01 cm$^{-1}$ instrumental resolution, as displayed on Figure 2 (A). Only one broad band is observed around 976 cm$^{-1}$ (full width half maximum FWHM = 2 cm$^{-1}$, 17 cm$^{-1}$ away from the solid phase band centre) that we assign to the $\nu_{as}$ mode, predicted to be 14 times more intense than the $\nu_s$ one.[44] Such a spectrum doesn't allow us to extract reliable structural data on (CH$_3$)ReO$_3$, which justifies the need for supersonic jet high resolution FTIR spectroscopy.

In a recent study of the urethane molecule (solid phase at room temperature, vapour pressure < 0.1 mbar at 300 K)[52], the Laboratoire de Dynamique, Interactions et Réactivité (LADIR) acquired an experimental know-how in producing rare-gas supersonic jets seeded with sublimated molecules. A glass container, filled here with MTO, and all the gas pipes all the way to the nozzle exit are heated up to ~350 K to prevent recondensation of the compound. The MTO vapour obtained is continuously seeded in a carrier gas to achieve backing pressure conditions compatible with a supersonic expansion. In practice, MTO/argon mixtures (backing pressure ~45 mbar) with an MTO molar dilution up to 13 % expand through a circular nozzle of 1-mm diameter. Jet-cooled MTO is finally probed by the 16-pass arrangement of the infrared beam of a Bruker IFS 120 HR interferometer equipped with a bandpass optical filter centred at 980 cm$^{-1}$ (FWHM = 100 cm$^{-1}$) and focussed on a HgCdTe photovoltaic detector (cut-off below 830 cm$^{-1}$). In our conditions of supersonic expansion, 15 g of crystalline MTO have been necessary to record the Fourier transform of 300 co-added interferograms at 0.005 cm$^{-1}$ instrumental resolution (0.0034 cm$^{-1}$ boxcar apodized). A liquid nitrogen trap positioned between the secondary and primary pumps enabled to retrieve ~50 % of the initial MTO quantity.

The FTIR jet spectrum of the $\nu_{as}$ band recorded at 0.005 cm$^{-1}$ instrumental resolution is shown in Figure 2 (B). The argon moderate backing pressures and MTO high dilution used for the supersonic expansion led to an efficient rovibrational cooling, as proved by the absence of unresolved broad structure due to hot bands. The spectrum is composed of two similar rovibrational contours centred at 975.98 cm$^{-1}$ and 976.60 cm$^{-1}$ respectively with a 10.5/6 intensity ratio. From the isotopic abundance of rhenium, 62.93 % for $^{187}$Re and 37.07 % for $^{185}$Re, we assign the lowest frequency band to the heaviest isotopologue, $i.e.$ $(CH_3)^{187}ReO_3$ and the highest one to $(CH_3)^{185}ReO_3$. Figure 3 (obs.) shows a zoom on the $^{187}$Re isotopologue band.

One first important conclusion of these experiments, obvious on Figure 2 and Figure 3 on which CO$_2$ laser lines are labeled, is that the $\nu_{as}$ band of MTO should have transitions that are compatible with the CO$_2$ laser spectral window.

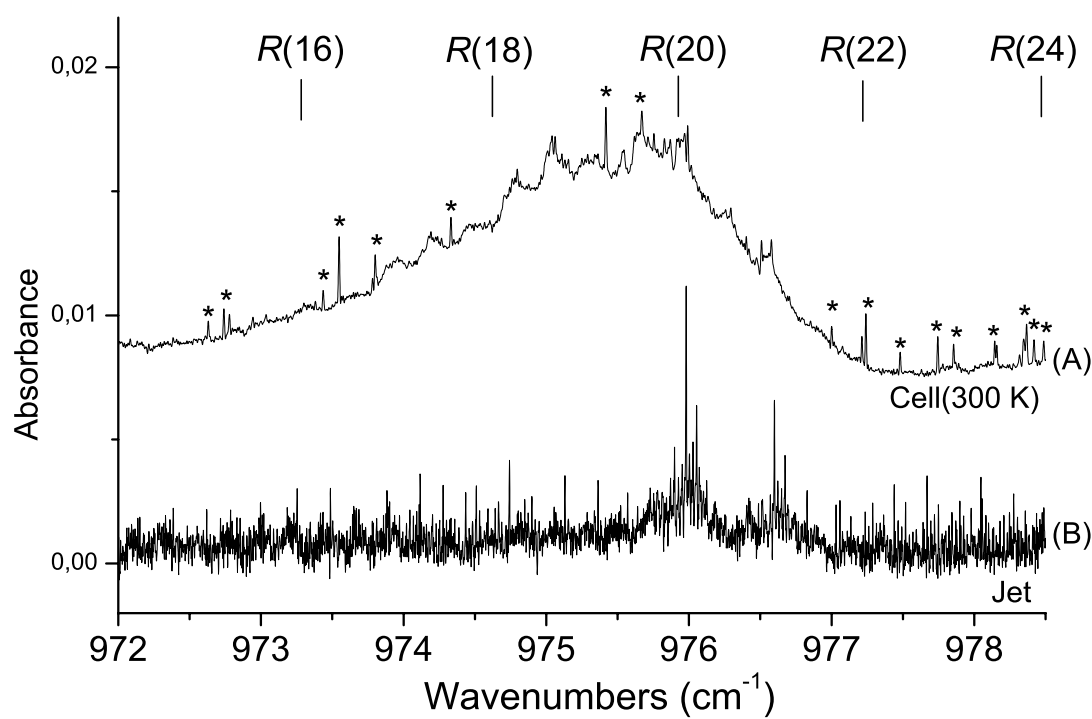

**Figure 2:** Room temperature cell (A) and jet-cooled (B) Fourier transform spectra of the $\nu_{as}$ band of MTO recorded at respective instrumental resolution 0.01 and 0.005 cm$^{-1}$. On the cell spectrum, narrow lines marked by an asterisk belong to rovibrational transitions of gas phase decomposition products of MTO. On the jet-cooled spectrum, the bands corresponding to the two $^{187}$Re and $^{185}$Re isotopic species are well resolved. $R(J)$ CO$_2$ laser lines are labeled.

## B. Analysis of the jet-cooled $\nu_{as}$ spectrum of MTO

As can be seen on Figure 3, each band contour of the $\nu_{as}$ perpendicular band[53] consists, near the band centre, in a series of strong $^{P}Q(J,K)$ ($\Delta K = -1$, low frequency side) and $^{R}Q(J,K)$ ($\Delta K = +1$, high frequency side) branches for which $K$ is fixed. Each branch is composed of a number of sub-bands of given $J$ with $J \geq K$.

The energy terms for the ground state rotational structure are given by the following expression:

$$E(J,K) = BJ(J+1) + (A-B)K^2 - D_{JJ}J^2(J+1)^2 - D_{JK}J(J+1)K^2 - D_{KK}K^4 \qquad (1)$$

Higher order terms were neglected. Only the first order terms of the diagonal part of the excited state vibration-rotation energy were used in our analysis:

$$E(\nu_{as},J,K,l) = \nu_{as} + B'J(J+1) + (A'-B')K^2 - 2\zeta A'Kl - D_{JJ}'J^2(J+1)^2 \\ - D_{JK}'J(J+1)K^2 - D_{KK}'K^4 \qquad (2)$$

where $\zeta$ is the first-order Coriolis constant for the degenerate vibration $\nu_{as}$.

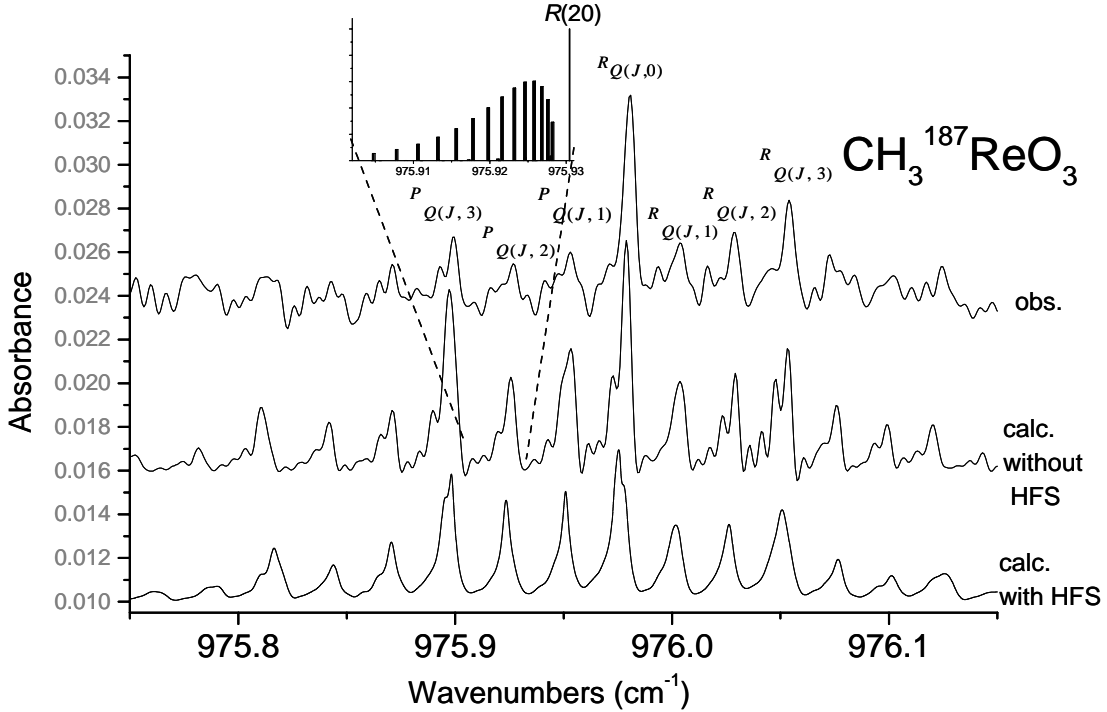

**Figure 3:** Comparison between the jet-cooled FTIR spectrum (obs.) of the $\nu_{as}$ band of $(CH_3)^{187}ReO_3$ in the $Q$ branch region and two simulations (see text) performed without (calc.without HFS) or with (calc.with HFS) the contribution of the hyperfine structure (HFS). The inset shows a simulated stick spectrum (before convolution with the boxcar response function) of the $^{P}Q(J,2)$ branch, composed of sub-bands of given $J$ with $J \geq K = 2$, without HFS. The $R(20)$ CO$_2$ laser line is labeled.

For a perpendicular band, transitions obey the rotational quantum number selection rules $\Delta J = 0, \pm 1$; $\Delta K = \pm 1$; $\Delta l = \pm 1$; $\Delta |K - l| = 0$. For the fundamental band, only the following combinations of parameters could be fitted: $\nu_{as} - \zeta A'$, $A' - A$, $A'(1 - \zeta)$. Meanwhile, by fixing the value of either $A$ or $\zeta$, independent values of the other molecular parameters can be determined.

Line intensities for such a symmetric top molecule perpendicular band can be approximated under the form:

$$I(J,K) = C A_{KJ} \nu g_{KJ} \exp[-E(J,K)h/kT_R] \qquad (3)$$

where $J$ and $K$ are the lower state quantum numbers, $A_{KJ}$ is the Hönl-London factor[53], $g_{KJ}$ the lower state spin-rotation statistical weight, $T_R$ is the rotational temperature, $\nu$ is the transition frequency, $C$ is a factor depending on the induced electric dipole moment and on the partition function, and $h$ and $k$ are the Planck and Boltzmann constants. The statistical weight $g_{KJ}$ is namely governed by the $2J+1$ rotational degeneracy factor and the rules of nuclear spin statistics for a $C_{3v}$ molecule with three identical atoms of nuclear spin 1/2 (H), outside the axis of symmetry. Taking into account the selection rules for a perpendicular band associated with a doubly degenerate upper state, the observed structure shows that the transitions from vibrational ground state rotational levels with $K = 3n$ are twice stronger than these associated to levels with $K = 3n \pm 1$ ($n$ is an integer).

For the analysis we followed a three-step strategy. We first roughly simulated the band contour of the, unresolved in $J$, $^PQ(J,K)$ and $^RQ(J,K)$ branches using formula (1) and (2), neglecting at that step the hyperfine structure (HFS). We used $A = 3849.81$ MHz (as estimated in section III) and other ground state parameters values obtained from microwave spectroscopy (see Table I), and took excited-state centrifugal distortion constants $D_{JJ}'$ and $D_{JK}'$ equal to those of the ground state. Then by manually adjusting $\nu_{as}$, $A'$, $B'$, $\zeta$ and $T_R$, we simply tried to reproduce the frequency maxima of the $Q$ branches (fourteen visible on Figure 3 for $^{187}$Re MTO). Each calculated transition was convolved with the boxcar response function at the instrumental resolution of 0.005 cm$^{-1}$ FWHM. The excited-state parameter set was obtained by manually minimizing the difference between experimental and simulated frequency maxima of the $Q$ branches. For $(CH_3)^{187}ReO_3$ our best agreement was obtained with $\nu_{as} = 975.968$ cm$^{-1}$, $A' = 0.12811$ cm$^{-1}$, $B' = 0.11554$ cm$^{-1}$ and $\zeta = -0.0027$. The rotational temperature $T_R$ was estimated to 10 ± 3K. Figure 3 (obs. and calc. without HFS) illustrates the good agreement, in the $Q$ branch region, between the jet FTIR spectrum of $(CH_3)^{187}ReO_3$ and the spectrum calculated with those parameters.

In a second step, we were able to assign 43 $^RR$ and 25 $^PP$ transitions of $(CH_3)^{187}ReO_3$ on the jet-cooled FTIR spectrum. The list of these assigned lines, their experimental and calculated frequencies, is available in the ancillary file.

For the less abundant $^{185}$Re MTO isotopologue, the same analysis led to the following set of parameters: $\nu_{as} = 976.588$ cm$^{-1}$, $A' = 0.12808$ cm$^{-1}$, $B' = 0.11555$ cm$^{-1}$ and $\zeta = -0.0027$. A few $^RR$ and $^PP$ transitions of $(CH_3)^{185}ReO_3$ were assigned, but a global fit could not be performed. These results are not reported here.

Up to now, the HFS has not been taken into account in our analysis of the antisymmetric Re=O vibrational mode. The strong quadrupole coupling strength $eQq > 700$ MHz (see Table I) of the rhenium, similar in magnitude to the rotational constants, is responsible for an important break-up of microwave rotational transitions into hyperfine components as observed in [45, 48] and will be responsible for large hyperfine splittings of the transitions simulated in the previous rovibrational analysis. This HFS cannot be resolved using FTIR jet spectroscopy but it is essential to fully include it in the rovibrational simulation for two reasons: (i) evaluate the HFS impact on the recorded FTIR spectra at 100 MHz (0.0034 cm$^{-1}$) instrumental resolution and on the corresponding rovibrational analysis, (ii) propose a list of transition frequencies of the $\nu_{as}$ band assigned in $J$, $K$ and $F$, in order to reproduce and interpret the resolved hyperfine structures observed at LPL with ultra-high resolution CO$_2$ laser spectroscopy techniques (see section V).

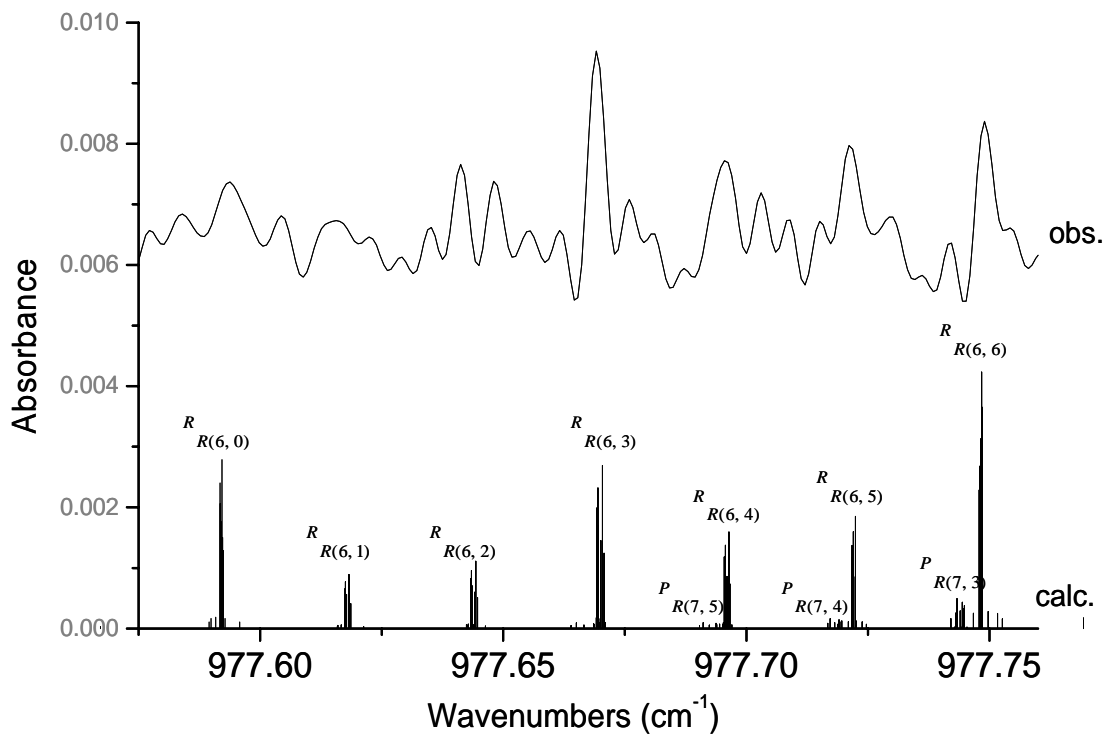

**Figure 4:** Comparison between observed jet-cooled FTIR and the calculated (with Set 2 of Table II) stick spectra of seven $^RR(6,K)$ assigned transitions (with $K$=0-6) of the $\nu_{as}$ band of (CH$_3$)$^{187}$ReO$_3$.

Thus, in a third step, we added the HFS with the help of the SPFIT/SPCAT suite of programs of H. Pickett[47], to the analysis of the $^{187}$Re MTO. We fitted the energy levels with a Hamiltonian containing the spin-rotation interaction terms $C_{aa}$ and $C_{bb}$, and more important, the rhenium quadrupole coupling constant $eQq$ in both the ground and excited states. For the $^{187}$Re isotopologue of MTO, we included in the fit: (i) the assigned rotational lines used in the microwave study of section III, (ii) the assigned $^PP(J,K)$ and $^RR(J,K)$ lines recorded by FTIR spectroscopy, and (iii) the HFS of the $^RQ(20,0)$ line recorded at ultra-high resolution at LPL, and the high resolved $^RQ(19,0)$ and $^PQ(12,1)$ lines recorded in a jet apparatus at LPL, as described in the next section, that could all be assigned over the course of this analysis (see section V for more details). The complete set of assigned lines used in this analysis is available in the ancillary file. The fit led to an rms value of (obs-cal) of 0.0017 cm$^{-1}$ (51 MHz) and the values obtained for the molecular parameters are presented in Table II together with their error bars. For this fit, centrifugal distortion constants (except $D_{KK}$) and the spin-rotation interaction $C_{aa}$ were fitted but constrained to the same value in the ground and excited states. At a first stage (Set 1 in Table II), we constrained the value of the *A* constant at 3849.81 MHz (value estimated in section III), and obtained $\zeta$ = -0.0011(3)<<1. At a second stage (Set 2 in Table II), we neglected the first-order Coriolis contribution, to be able to fit both effective *A* and *A'* rotational constants. The *A* value stays within the predicted uncertainty (section III). Both sets of parameters are equivalent.

Figure 3 illustrates the good agreement, in the *Q* branch region, between the jet FTIR spectrum of $(CH_3)^{187}ReO_3$ and the spectrum calculated with Set 2 of Table II. Compared to the rough simulation of the first step, adding the HFS contribution does not practically change the position of the maxima of the *Q* branches but has for consequence to smooth their overall line shape due to the unresolved structure of the rovibrational transitions. Figure 4 displays a comparison between jet-cooled and calculated (with Set 2 of Table II) stick spectra of seven $^RR(6,K)$ assigned transitions (K from 0 to 6) of the $\nu_{as}$ band of $(CH_3)^{187}ReO_3$. The HFS contribution is clearly illustrated on this figure. The order of magnitude of HFS splitting is ~100 MHz (~$3\times10^{-3}$ cm$^{-1}$) for low *J* values. It decreases when *J* increases, and increases with *K*.

**Table II :** Ground- and excited-state ($\nu_{as}$ = 1) parameters of the isotopologue $CH_3^{187}ReO_3$. Microwave (section III) and Infrared data, both from FTIR (section IV) and high resolution $CO_2$ laser absorption experiments (section V), were used. The listed uncertainties are 1σ. Both sets of values are equivalent (see text).

|  | Set 1 | Set 2 |
|---|---|---|
| *A* (MHz) | 3849.81 [a] | 3854.01(1.27) |
| *B* (MHz) | 3466.96481(39) | 3466.96481(39) |
| $D_J$ (kHz) | 0.705(50) [b] | 0.705(50) [b] |
| $D_{JK}$ (kHz) | 2.208(118) [b] | 2.208(118) [b] |
| *eQq* (MHz) | 716.54005(192) | 716.54005(192) |
| $C_{aa}$ (kHz) | -52.22(37) [b] | -52.22(37) [b] |
| $C_{bb}$ (kHz) | -51.464(92) | -51.464(92) |
| $\nu_{as}$ (cm$^{-1}$) | 975.9665(3) | 975.9667(3) |
| *A'* (MHz) | 3847.14(34) | 3851.35(1.12) |
| *B'* (MHz) | 3463.4362(224) | 3463.4362(224) |
| $\zeta$ | -0.0011(4) | 0.0 [a] |
| $D_J$' (kHz) | 0.705(50) [b] | 0.705(50) [b] |
| $D_{JK}$' (kHz) | 2.208(118) [b] | 2.208(118) [b] |
| *eQq*' (MHz) | 694.779(44) | 694.779(44) |
| $C_{aa}$' (kHz) | -52.22(37) [b] | -52.22(37) [b] |
| $C_{bb}$' (kHz) | -53.005(149) | -53.005(149) |

(a) Fixed value
(b) Fitted and constrained to the corresponding ground state/excited state value

# V. Ultra-high resolution absorption spectroscopy of MTO

In the context of molecular PV observation, the highest resolution spectroscopy techniques are needed and an experimental set-up based on the powerful method of Doppler-free two-photon Ramsey fringes in a supersonic molecular beam was proposed.[25-28] It was thus necessary to check whether a complex molecule such as MTO which has spectroscopic properties similar to the chiral derivatives that we consider for such a test, is suitable for this kind of spectroscopy and optimize the supersonic beam apparatus accordingly. The first steps in this direction are exposed in the following paragraphs.

## A. The spectrometer

The ultra-high resolution laser spectrometer used for the experiment dedicated to the PV observation is based on the combination of two $CO_2$ lasers which can operate in the 8-12 µm range (already described elsewhere[29, 30]). Frequency stability and tunability are important issues for this experiment. In order to probe lines of the $\nu_{as}$ band of MTO, found to be centred around 976 cm$^{-1}$, the two lasers are set on one same laser line of the $CO_2$ 10.2-µm $R$ branch. The first is locked to an $OsO_4$ rovibrational line. The frequency stabilization scheme is described in [54]: a sideband generated with a tunable electro-optic modulator (EOM) is stabilized on an $OsO_4$ saturated absorption line (FWHM ~20 kHz) detected in transmission of a 1.6-m long Fabry-Perot cavity. The laser spectral width measured from the Allan deviation of the beat note between two independent lasers shows a typical frequency instability below 1 Hz after 100 s.[54] For data collection and averaging (see section C), frequency stability is however also required over longer times of a few tens of minutes, after which the typical frequency instability is ~100 Hz.[54] This system can achieve a frequency accuracy better than 100 Hz.[54]

The second laser, whose beam is used to probe MTO, is phase locked to the first with a tunable radio frequency (RF) offset. The first laser sets the frequency stability and accuracy, while the second enables frequency tunability under computer control of the synthesizer delivering the RF offset. In these conditions, tunability is however limited to the laser gain curve width, *i.e.* ~80 MHz. By allowing, or not, the second laser beam to be frequency shifted by up to two acousto-optic modulators (AOMs), tunability around each $CO_2$ line could be increased to a spectral window of [-280 MHz, +280 MHz]. Both AOMs, of respective fixed frequencies 40 and 80 MHz, were set up in double pass, either on the +1 or -1 diffraction order.

## B. Saturated absorption spectroscopy in a cell

In this section, we describe saturated absorption spectroscopy of MTO at 300 K in a cell.[25] These experiments constitute a first step towards jet spectroscopy (much more molecule-consuming). Less than 1 g of MTO was enough to perform this study. They were carried out to confirm results of section IV analysis predicting that lines were observable with our $CO_2$ laser spectrometer, and to try to identify predicted rovibrational transitions.

A 58-cm long and 45-mm diameter cylindrical glass cell, ended with anti-reflective coated ZnSe windows, was filled with a vapour of MTO at a pressure of about 10$^{-3}$ mbar much below the vapour pressure of a few 10$^{-1}$ mbar at room temperature. The laser beam was sent in the cell (pump beam), retroreflected with a simple mirror (probe beam) and focused on an HgCdTe photodetector. The power ratio between pump and probe beam could be controlled by placing optical attenuators between the mirror and the cell. The beam waist inside the cell was 5 to 10 mm. To reduce the effect of laser amplitude noise, the saturated absorption signal is detected after frequency modulation at 5 kHz. The modulation is applied on piezoelectric transducers that control the length of the laser cavity. A locking amplifier enables second harmonic detection in order to strongly flatten the baseline induced by the laser gain curve. Spectra are recorded by scanning the laser frequency through computer control of the phase-lock loop RF offset.

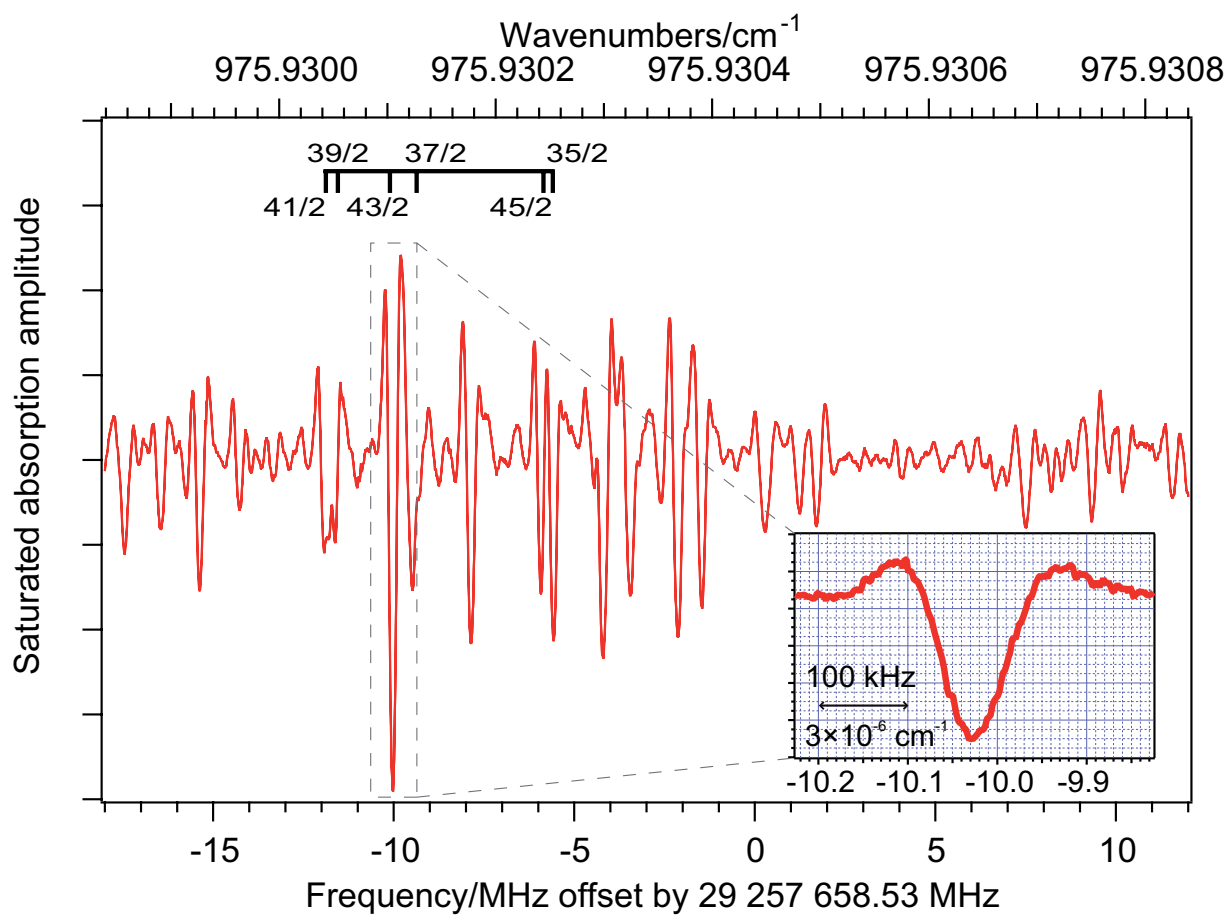

**Figure 5:** Saturated absorption spectrum of MTO in a cell at 300 K (experimental conditions: 3000 points, 100 ms integration time per point, average of 2 back-and-forth sweeps, 200 kHz of modulation depth (modulation index of 40), 0.002 mbar of MTO pressure, pump (resp. probe) beam power of 95 µW (resp. 12 µW)). The offset corresponds to the $R(20)$ $CO_2$ laser line frequency. The inset is a zoom on a single intense line (experimental conditions: 200 points, 100 ms integration time per point, single back-and-forth sweep, 50 kHz of modulation depth (modulation index of 10), 0.002 mbar of MTO pressure, pump (resp. probe) beam power of 24 µW (resp. 3 µW)). Assigned $\Delta F=0$ six most intense hyperfine components associated with the $^RQ(20,0)$ line of $^{187}$Re MTO are labelled with their $F$ quantum number.

Saturated absorption spectra were recorded with MTO pressures typically ranging from $10^{-4}$ to $10^{-2}$ mbar, pump beam power ranging from 5 µW to 2.5 mW and pump to probe beam power ratio ranging from 1 to 8. The first laser is locked on an $R(67)$ line of $^{192}$OsO$_4$.[55] Figure 5 displays a spectrum recorded over 30 MHz around the $R(20)$ $CO_2$ laser line (origin of the frequency axis), where we expect to find several MTO lines from the previous rovibrational study (section IV). The hyperfine structure of MTO is responsible for such a dense spectrum. The inset zooms on a single intense line, recorded over 400 kHz. This line is well fitted by the 2$^{nd}$ derivative of a Lorentzian of half width at half maximum (HWHM) of 94 kHz, and is centred at 29 257 648.504(2) MHz (975.93010508(4) cm$^{-1}$). By reducing pressure (~10$^{-4}$ mbar), modulation depth (30 kHz) and laser intensity (~µW), HWHM line widths smaller than 50 kHz (~10$^{-6}$ cm$^{-1}$) were obtained. Such a residual line width can not be explained only by transit time broadening (~5 kHz) and residual Doppler broadening induced by an angle between the two counter-propagating beams (<20 kHz), but is fully compatible with an unresolved magnetic hyperfine structure due to the spins of Re and H nuclei. This particular line was assigned as the $\Delta F$=0, $F$=43/2 quadrupolar component of the $^RQ(20,0)$ rovibrational line (see below).

Ultra-high resolved dense signals such as the one on Figure 5, were observed in the probed [-200 MHz, +270 MHz] spectral window around the $R(20)$ $CO_2$ laser line. Less dense spectra of smaller amplitude were observed in spectral windows of ~50 MHz around $R(18)$ and $R(18)$-160 MHz, $R(22)$ and $R(22)$ -160 MHz, $R(24)$ and $R(24)$ -160 MHz (with the first laser respectively locked on the $R(59)A_1^3(-)$ line of $^{190}OsO_4$, and $R(74)A_1^0(-)$ and $R(80)A_1^2(-)$ line of $^{192}OsO_4$ [55]). Those observations correlated rather well with the conclusions of our analyses of section III and IV. From the set of parameters obtained in the analysis including HFS, we simulated a 300 K spectrum. In the spectrum displayed on Figure 5, we managed to identify the $\Delta F$=0 six most intense hyperfine components associated with the $^RQ(20,0)$ line of $^{187}$Re MTO, and probably components of either higher $J$ lines or $^{185}$Re MTO unidentified signals. The measured frequencies of the six identified hyperfine components of the $^RQ(20,0)$ line, which are labelled with their $F$ quantum number on Figure 5, were actually included in the analysis with HFS, as mentioned in section IV.B, with a 10 kHz experimental uncertainty. Their assigned frequencies are available in the ancillary file with the complete set of assigned lines used in section IV.B analysis. Additionally, a list of $^{187}$Re MTO simulated (from Set 2 of Table II) lines that are potentially observable in the 560 MHz spectral region that can be explored around the $R(18)$, $R(20)$, $R(22)$ and $R(24)$ laser lines is given in Table III. The simulated frequency uncertainties $\sigma_f$ given by the fitting procedure were taken into account to build up this table. We noticed that $\sigma_f$ roughly follows $\sigma_f$ ~ 0.016 $J^2$+0.328 $K^2$ MHz (with however lower $\sigma_f$ for $J$ ~ 20, as a consequence of the assignment of the $^RQ(20,0)$ hyperfine components with a very good experimental uncertainty). In contrast to the signal observed around $R(18)$ and $R(20)$ laser lines, which can potentially originate from lots of lines, the situation is much simpler around $R(22)$ which is expected to coincide with only one simulated rovibrational line, $^RR(4,3)$. No $^{187}$Re MTO line is expected around $R(24)$. However the observed signal is compatible with the simulated $^RR(7,1)$ line of $^{185}$Re MTO as deduced from the rougher analysis of this isotopologue (section IV). Note that we expect $^{185}$Re MTO to also contribute to the signal observed around the other $CO_2$ laser lines.

Higher-order terms, namely the non-diagonal contributions from the l-type doubling interactions, should be included in the model to fully reproduce the ultra high-resolution spectrum. It would then be possible to propose new assignments with estimated uncertainties. In the near future, the identification in the saturation spectra of more simulated lines (in particular the two isolated lines expected around $R(22)$ and $R(24)$) should lead to a refined set of molecular parameters. This work is currently under progress. However, the resolution is not good enough for a complete study of magnetic hyperfine interactions. This could only be performed by using our 18-m long absorption cell[56] or by observing Ramsey fringes on a molecular beam.[29]

**Table III:** Calculated lines of $^{187}$Re MTO potentially observable in the 560 MHz spectral region around $R(18)$, $R(20)$, $R(22)$ and $R(24)$ $CO_2$ laser lines. The $(J,K)$ quantum numbers are listed in the third column. Already experimentally assigned lines are indicated in bold

| | | |
|---|---|---|
| $R(18)$ | $^PQ$ | (45,45), (46,45), (47,44), (47,45), (47,46), (48,44), (48,46), (48,45), (49,43), (49,44), (49,45), (50,43), (50,44), (50,45), (51,42), (51,43), (51,44), (52,42), (52,43), (52,44), (52,43), (53,41), (53,42), (53,43), (54,41), (54,42), (55,40), (55,41), (55,42), (56,40), (56,41), (57,39), (57,40), (57,41), (58,40), (59,38), (59,39), (59,40), (60,38), (60,39), (61,37), (61,38), (62,36), (62,37), (62,38), (62,36), (63,37), (64,35), (64,36), (65,35), (65,36), (66,34), (66,35), (67,33), (67,34), (68,34), (68,33), (69,32), (69,33), (70,31), (70,32), (71,31), (71,32), (72,30), (72,31), (73,30), (74,30), (75,28), (75,29), (76,28), (77,27), (79,25), (79,26), (80,25), (84,22), (85,21), (73,29), (74,29), (76,27), (77,26), (78,26), (80,24), (81,23), (81,24), (82,23), (83,22), (84,21), (85,20), (86,19), (86,20), (87,19), (88,18), (89,17), (90,16), (91,15), (91,16), (92,14), (92,15), (94,12), (95,12), (98,9) |
| | $^RP$ | (6,1) |
| $R(20)$ | $^PQ$ | (2,2), (3,2), (4,2), (5,2), (6,2), (7,2), (11,1), **(12,1)** [a], (13,1), (14,1), (15,1), (16,1) |
| | $^RQ$ | (18,0), **(19,0)** [a], **(20,0)** [b], (21,0), (24,1), (25,1), (26,1), (28,2), (29,2), (30,2), (31,3), (32,3), (33,3), (35,4), (36,4), (39,5), (38,5), (41,6), (42,6), (43,7), (44,7), (46,8), (47,8), (48,9), (49,9), (50,10), (51,10), (52,11), (53,11), (54,12), (55,12), (56,13), (57,13), (58,14), (59,14), (60,15), (61,15), (62,16), (63,16), (63,17), (64,17), (65,18), (66,18), (67,19), (68,19), (68,20), (69,20), (70,21), (71,21), (71,22), (72,22), (73,23), (74,23), (74,24), (75,24), (76,25), (77,25), (77,26), (78,26), (78,27), (79,27), (80,28), (81,28), (81,29), (82,29), (82,30), (83,30), (84,30), (84,31), (85,31), (85,32), (86,32), (86,33), (87,33), (87,34), (88,34), (89,34), (89,35), (90,35), (90,36), (91,36), (93,39), (94,39), (95,39) |
| $R(22)$ | $^RR$ | (4,3) |
| $R(24)$ | | Ø |

(a) assigned lines in the jet-cooled linear absorption spectrum of Figure 6
(b) assigned line in the 300 K saturated absorption spectrum of Figure 5

## C. Molecular jet spectroscopy

In this section, we detail the first high resolution but still Doppler broadened (~MHz line width) spectra of jet-cooled MTO. They were obtained on the set-up currently under development for molecular PV observation, and constitute a significant advance towards the ultimate test, likely to be performed on chiral derivatives of MTO.

The molecular jet apparatus is 3-m long with two vacuum chambers pumped by diffusion pumps. Supersonic expansion occurs through a circular nozzle in a first chamber (pressure of $10^{-5}$-$10^{-4}$ mbar under working conditions) separated from the second one (pressure of $10^{-6}$-$10^{-5}$ mbar under working conditions) by a skimmer. The nozzle-to-skimmer distance is adjustable between 2 and 25 mm. To produce a supersonic jet of MTO, our strategy (like in section IV) consisted in heating MTO contained in a reservoir (stainless steel) and seeding its vapour in a carrier gas (helium). Extensive studies of the MTO-seeded jet characteristics (MTO dilution and flux, longitudinal velocity, translational temperature,…) as a function of reservoir temperature, backing He pressure, nozzle-to-skimmer distance, nozzle diameter (ranging from 50 to 300 µm) and skimmer diameter (ranging from 0.7 to 2 mm) have been performed via time-of-flight experiments[57], and will be detailed elsewhere.

The experimental set-up for linear absorption spectroscopy of a jet of MTO is as follows[58]: after having crossed the molecular jet perpendicularly, the laser beam was retroreflected with a mirror (double-pass configuration), and its absorption was monitored as a function of laser frequency. A slotted disk was used to chop the molecular jet at ~750 Hz, thus inducing a modulation of the absorption, then detected through a lock-in amplifier. The line shape is Doppler broadened owing to the molecular jet divergence, but it is complicated by the fact that the modulation phase is not uniform as the molecular jet diameter at the chopper is not negligible compared to the slots' size.[58]

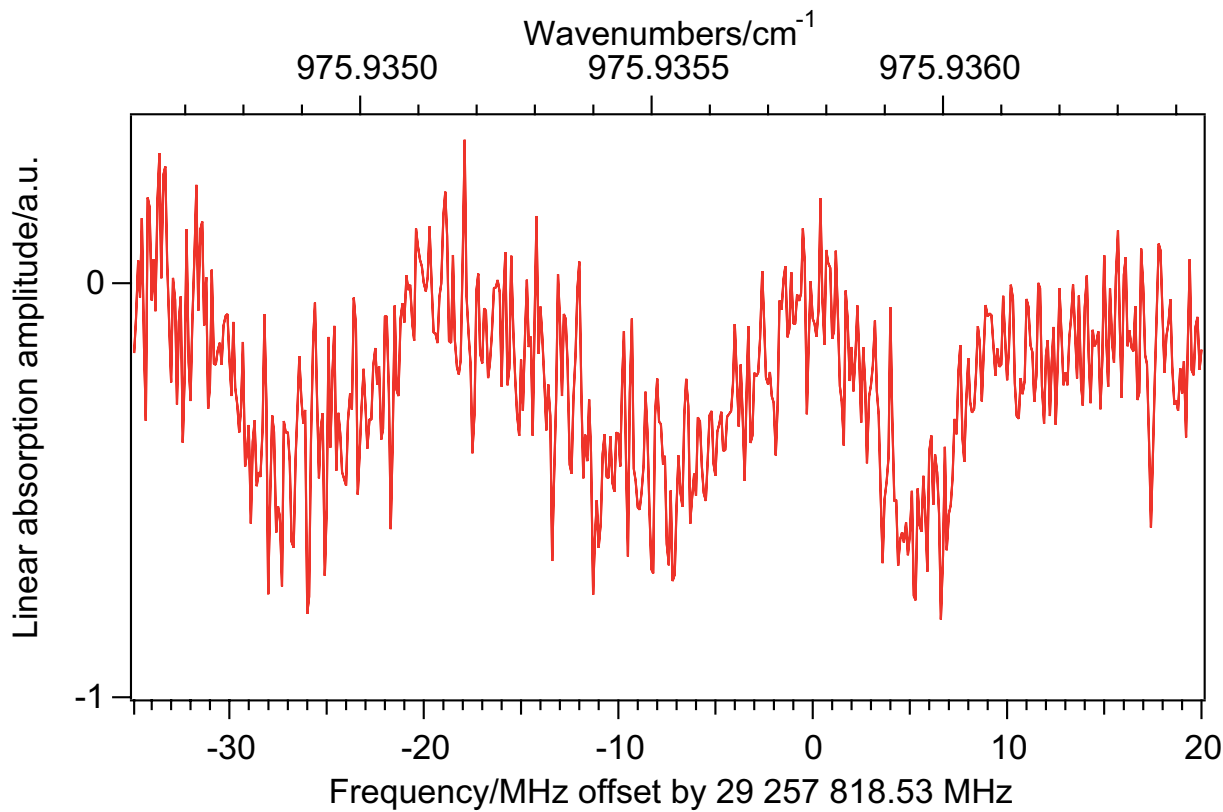

**Figure 6:** Linear absorption spectroscopy in a supersonic jet seeded with MTO. The origin of the abscissa axis is the $R(20)$ $CO_2$ laser line frequency + 160 MHz. Experimental conditions: 550 points, 100 ms integration time per point, a total integration time of ~20 min after back-and-forth averaging sweeps, laser beam power of ~2 µW (in single pass), MTO reservoir temperature of ~350 K, He backing pressure of ~150 mbar, nozzle (resp. skimmer) diameter of 200 µm (resp. 2 mm), nozzle-to-skimmer distance of 7.5 mm, translational temperature of ~7 K, longitudinal speed of ~850 m.s$^{-1}$, ~7% MTO dilution, equivalent MTO partial pressure of ~10 mbar in the reservoir, flux of $10^{18}$ molecules of MTO per second.

Figure 6 displays a spectrum centred +160 MHz away from the $R(20)$ $CO_2$ laser line, recorded over 55 MHz. The jet experimental conditions, some deduced from time-of-flight measurements, are given in the caption. The noise intensity is limited by the laser technical noise proportional to the power. Thus to maximise the signal-to noise ratio, the single pass laser power was chosen to be ~2 µW, the smallest possible to be just above the photodetector noise. On Figure 6, three Doppler broadened lines (between 2 and 5 MHz HWHM) respectively centred on (i) 29 257 792.2(3) MHz (975.934899(7) cm$^{-1}$), (ii) 29 257 810.2(3)

MHz (975.935499(8) cm$^{-1}$) and (iii) 29 257 824.0(2) MHz (975.935959(4) cm$^{-1}$) can be distinguished. Following the analysis of section IV, line (i) and (ii) were respectively assigned as $^{R}Q(19,0)$ and $^{P}Q(12,1)$ of the $^{187}$Re isotopologue and were actually included in the analysis with HFS (with an experimental uncertainty of a few MHz, equal to the line width) performed in section IV.B as mentioned previously. Line iii) is tentatively assigned as $^{R}Q(28,2)$ or $^{R}Q(24,1)$ of the $^{187}$Re isotopologue (but higher-order effects should be included in the model). The signal-to-noise ratio appears to be rather poor for ~20 min total integration time, but it constitutes an important step towards the observation of molecular PV. To our knowledge, this is the first ultra-high resolution (~1 MHz line width) molecular jet absorption spectrum, obtained on such a complex organometallic molecule.

The next step is to optimize the set-up and improve the signal to noise ratio by a factor of ~10, to be able: (i) to carry out a faster screening in the accessible spectral window and point more lines; (ii) to allow us to reduce the nozzle diameter and the molecule consumption accordingly (~2 g/h currently). For that purpose, the design of a multi-pass cell compatible with jet spectroscopy is under progress.[59]

## VI. Conclusion

In this paper a detailed spectroscopic study of the antisymmetric Re=O stretching mode of MTO has been undertaken, in the context of molecular PV observation. MTO is an ideal test molecule, as the achiral parent molecule of promising chiral candidates for the PV experiment.[25] The combined analysis of FTMW and FTIR jet-cooled $^{187}$Re MTO spectra enabled to assign 28 observed rotational lines in $J$, $K$ and $F$, and 68 observed rovibrational transitions in $J$ and $K$, and led to a list of calculated transition frequencies accessible to the CO$_2$ laser. First high resolution CO$_2$ laser absorption spectra of MTO have been recorded both in a cell (~100 kHz line width) and in a molecular jet (~1 MHz line width). A first relevant assignment of fully resolved hyperfine transitions for achiral MTO was proposed which, together with the FTMW and FTIR spectra analysis contributed to find a set of spectroscopic parameters in the ground and first excited state, including hyperfine structure constants. High resolution spectra of jet-cooled MTO have been observed, on the set-up being developed for the observation of molecular PV, which constitutes a major step towards the targeted objective.

One major aim of this work is, for a molecule for which very little spectroscopic data is available, to build up a whole procedure to be followed once the best candidate for a PV experiment will be synthesized. This work shows the feasibility and efficiency of the method which was a real concern before this study. These results highlight that several spectroscopic tools in very different spectral regions, with diverse experimental set-ups and various resolutions are essential to obtain a precise determination of ground and excited state parameters. In the future, the developed procedure will enable us to identify the most intense lines of chiral molecules accessible to our spectrometer, and to select the best candidate line to be studied for PV observation.

## Acknowledgements

This work was supported by ANR NCPMOL (Contract grant number : ANR-05-BLAN-0091), Ministère de la Recherche et de l'Enseignement Supérieur, CNRS, Région Bretagne, Rennes Métropole. O. Lopez, is gratefully acknowledged for his contribution to this project. We acknowledge valuable discussions with Trond Saue and Radovan Bast.